\documentclass[a4paper, orivec, runningheads]{llncs}
\usepackage{floatrow}
\newfloatcommand{capbtabbox}{table}[][\FBwidth]
\usepackage{tablefootnote}
\usepackage{graphicx}
\usepackage{amsmath}
\usepackage{array}
\usepackage{makecell, multirow, booktabs}
\usepackage{bm}
\usepackage{float}
\usepackage{url}
\usepackage{makeidx}  
\usepackage{graphicx}
\usepackage{subfigure}
\usepackage{amsmath}
\usepackage{amssymb}
\usepackage{verbatim}
\usepackage{algorithm}
\usepackage{algorithmic}
\usepackage{booktabs}
\usepackage{multirow}
\usepackage[misc]{ifsym} 
\usepackage{pbox} 
\usepackage{cite}
\usepackage{wrapfig}
\usepackage{threeparttable}

\usepackage{appendix}
\usepackage[colorlinks,
            linkcolor=blue,
            anchorcolor=blue,
            citecolor=blue
            ]{hyperref}
\usepackage{txfonts}
\begin{document}
\title{Rethinking Abdominal Organ Segmentation (RAOS) in the clinical scenario: A robustness evaluation benchmark with challenging cases}
\titlerunning{RAOS}
\author{Paper ID: 1633}
\institute{}
\authorrunning{X. Luo, Z. Li, S. Zhang, W. Liao and G. Wang} %
\author{Xiangde Luo\inst{1,2}
\and Zihan Li\inst{3} 
\and Shaoting Zhang\inst{1,2}
\and Wenjun Liao\inst{1}
\and Guotai Wang\inst{1,2}$^{(\textrm{\Letter})}$}

\institute{
$^1$University of Electronic Science and Technology of China, Chengdu, China\\
$^2$Shanghai AI Lab, Shanghai, China\\
$^3$University of Washington\\
Emails: (\email{luoxd1996@gmail.com}, \email{guotai.wang@uestc.edu.cn})\\
}
\maketitle
\footnote{X. Luo and Z. Li contributed equally to this work.}
\vspace{-8mm}

\begin{abstract}
Deep learning has enabled great strides in abdominal multi-organ segmentation, even surpassing junior oncologists on common cases or organs. However, robustness on corner cases and complex organs remains a challenging open problem for clinical adoption. To investigate model robustness, we collected and annotated the RAOS dataset comprising 413 CT scans ($\sim$80k 2D images, $\sim$8k 3D organ annotations) from 413 patients each with 17 (female) or 19 (male) labelled organs, manually delineated by oncologists. We grouped scans based on clinical information into 1) diagnosis/radiotherapy (317 volumes), 2) partial excision without the whole organ missing (22 volumes), and 3) excision with the whole organ missing (74 volumes). RAOS provides a potential benchmark for evaluating model robustness including organ hallucination. It also includes some organs that can be very hard to access on public datasets like the rectum, colon, intestine, prostate and seminal vesicles. We benchmarked several state-of-the-art methods in these three clinical groups to evaluate performance and robustness. We also assessed cross-generalization between RAOS and three public datasets. This dataset and comprehensive analysis establish a potential baseline for future robustness research: \url{https://github.com/Luoxd1996/RAOS}.

\end{abstract}

\section{Introduction}
Accurate and robust organ segmentation is irreplaceable in abdominal malignancy diagnosis, treatment, and follow-up. Especially in radiation therapy, inaccurate organ segmentations might lead to dose miscalculations and further bring under-treatment for malignancy tumors and unexpected side effects for normal organs~\cite{liao2023comprehensive}. Previously, the organ-at-risk contours were implemented by senior oncologists manually. Recently, many deep learning-based methods achieved promising segmentation in abdominal organ segmentation tasks~\cite{isensee2021nnu,lee20223d}. These efforts show the potential to boost the clinical delineation flow and reduce the delineation burden and time. However, their robustness and generalization on real clinically challenging cases still have not been investigated, limiting their clinical applications. The potential reason is lacking of public datasets and benchmarks as few recent datasets have considered real clinical scenarios.

\par Recently, with the efforts of the whole community~\cite{ji2022amos,ma2021abdomenct,kavur2021chaos,gibson2018automatic,luo2022word,antonelli2022medical,wasserthal2022totalsegmentator}, several datasets were developed for the model development and performance evaluation. These datasets show that deep learning can perform well on some common abdominal organs, such as the liver, spleen, pancreas, kidney, and stomach. But for small and complex anatomical organs (e.g., gallbladder, duodenum, and adrenal), it's still a challenging problem to achieve promising segmentation. Although these datasets improve the abdominal organ segmentation research, there are still some abdominal organs not being comprehensively investigated, such as the colon and intestine not in AMOS~\cite{ji2022amos}. In addition, most of these datasets~\cite{luo2022word,ma2021abdomenct} mainly focus on investigating the robustness and generalization caused by the shift in intensity distribution due to imaging protocols, scanners, or others. For more challenging clinical phenomenons, like patients with partial excision or whole excision surgery, there are few studies evaluating the robustness and generalization of models on these corner cases. And there has not been still a benchmark for the models' robustness and generalization evaluation on corner cases.

\par Based on the above observations, we present RAOS, a purely manually labelled whole abdominal organ segmentation dataset with different imaging, diseases, and treatment strategies, including 19 common organ annotations required in abdominal radiotherapy planning. Different from previous datasets, RAOS has the following attributes: 1)  comprehensive annotations, RAOS consists of 413 Computerized Tomography (CT) scans with over 84k slices, 7.4K 3D organs, and 19 types of abdominal organs, which may be the largest and most comprehensive manually labelled abdominal organ segmentation dataset. In addition, the RAOS provided professional annotations on some organs not shown in previous public datasets (such as the rectum, colon, intestine, prostate, seminal vesicles, and head of femur), which can improve the diversity of the public abdominal datasets in both data distribution and organ classes. 2) clinically corner cases, RAOS acquires images from patients suffered from different cancers and with different treatment strategies (immunotherapy, radiotherapy, chemotherapy, partial excision or whole excision surgery or combinations of them). Due to the surgery intervention, the normal anatomy is braked and introduces a new domain shift in the anatomy itself and further requires higher robustness and generalization of models. According to the two characteristics, RAOS shows the potential ability to be an evaluation benchmark for clinical corner cases.

\par With the efforts of data collection and manual annotation, we attempt to propose a new dataset for robustness and generalization evaluation on clinical corner cases and build a benchmark on several recent medical image segmentation methods. Specifically, we found that these methods can achieve encouraging results on the without surgery subset, most of them DSC and NSD are larger than 80\%. However, these methods' performance dropped significantly on two with surgery subsets, specifically on the surgery with organ missing subset, the DSC dropped near 6\%. These benchmarking results demonstrate that state-of-the-art methods can not produce clinically applicable segmentation on corner cases and with the clinical challenges increasing, the performance of these methods dropped significantly. The main contributions can be roughly summarized as two-fold: 1) We built a large-scale abdominal organ segmentation dataset with clinically challenging cases and 19 annotated organs, which is a more clinical and challenging dataset than previous, enabling the robustness evaluation in clinically challenging cases; 2) We established the RAOS benchmark by investigating the state-of-the-art (SOTA) methods' performance on clinically challenging cases and introducing the organ hallucination ratio to measure the SOTA methods' robustness on resection patients.
\section{Related Work}
\subsection{Abdominal Organ Segmentation}
\textbf{\textit{Datasets and Benchmarks.}} Thanks to the community's efforts, several abdominal organ segmentation datasets have been built and released. The popular datasets include BTCV~\cite{landman2015miccai}, MSD~\cite{antonelli2022medical}, AMOS~\cite{ji2022amos}, WORD~\cite{luo2022word} and Abdomenct-1k~\cite{ma2021abdomenct} datasets. The BTCV dataset~\cite{landman2015miccai} has 30 individuals with abdominal portal venous contrast enhancement CT scans with 13 organ annotations. The main goal of the MSD dataset~\cite{antonelli2022medical} is to improve algorithms so that they can be applied to various tasks, rather than focusing on achieving the best performance in all 10 tasks. The AMOS dataset~\cite{ji2022amos} has 500 CT scans and 100 MRI scans from different medical centers, equipment manufacturers, imaging techniques, disease types, and stages where each scan has 15 organ annotations. The WORD dataset~\cite{luo2022word} has 150 whole abdominal region CT volumes with 16 organ annotations. The AbdomenCT-1K dataset~\cite{ma2021abdomenct} has 1000+ CT scans from 12 medical centers with 4 organ annotations \\ \textbf{\textit{State-of-the-art methods.}} There are several segmentation methods~\cite{gibson2018automatic, isensee2021nnu,lee20223d,lee2023scaling,zhou2023nnformer,tang2022self,wang2021transbts,hatamizadeh2022unetr,li2022tfcns} to address the issue of the abdominal organ segmentation. Among them, nnU-Net~\cite{isensee2021nnu} is a segmentation method based on deep learning that can adjust to new tasks by customizing itself, such as preprocessing, network structure, training, and post-processing. Inspired by vision transformers, Tang et al.~\cite{tang2022self} introduce Swin UNETR for abdominal organ and tumor segmentation. It converts multi-modal input into 1D embeddings, which are encoded by a hierarchical Swin transformer. Hybrid methods succeed due to their wide non-local self-attention coverage and a large number of model parameters, as per researchers. So Lee et al.~\cite{lee20223d} present a compact 3D UX-Net, a ConvNet-based approach, that includes ConvNet modules for precise volumetric segmentation.

\subsection{Robustness on Cases With Organ Resection }
Despite the existence of datasets have been built for multi-organ segmentation, such as Totalsegmentator\cite{wasserthal2022totalsegmentator}, WORD\cite{luo2022word}, AMOS~\cite{ji2022amos} and Abdomenct-1k~\cite{ma2021abdomenct} datasets.
Furthermore, there is a growing interest in developing segmentation methods that cater to specific clinical scenarios. These methods include multimodal segmentation~\cite{guo2019deep,zhang2021modality,li2023lvit,shan2023coarse}, which involves utilizing data from various modalities to accurately segment desired objects, and scribble-supervised segmentation~\cite{luo2022scribble,li2023scribblevc}, which involves using annotated scribbles to guide the segmentation process. However, these methods are only inspired by various clinical scenarios of data or annotation acquisitions, without taking into account difficult clinical situations, like corner case scenarios. The primary concern we have is the phenomenon of organ segmentation illusion that occurs after the completion of organ resection surgery. Rickmann et al.~\cite{rickmann2023halos} conducted preliminary research on hallucination and found that sophisticated segmentation models often generate hallucinations of organs following organ removal. These hallucinations are inaccurate predictions of organs that cannot be rectified through oversampling or post-processing. Our research goes beyond HALOS~\cite{rickmann2023halos} by offering a more thorough examination of organ deficiencies. Our dataset covers not only the lack of gallbladders but also the absence of 7 other organs, including the prostate. Unlike HALOS~\cite{rickmann2023halos}, we conducted a detailed quantitative analysis of the findings, revealing that hallucinations resulting from organ loss are widespread and not specific to any particular organ. Furthermore, our study compared the challenging clinical segmentation scenario with traditional medical image segmentation scenarios, and the research demonstrated that learning from difficult segmentation samples can enhance the model's robustness and generalizability.
\begin{figure*}[t]
    \centering
\includegraphics[width=0.98\textwidth]{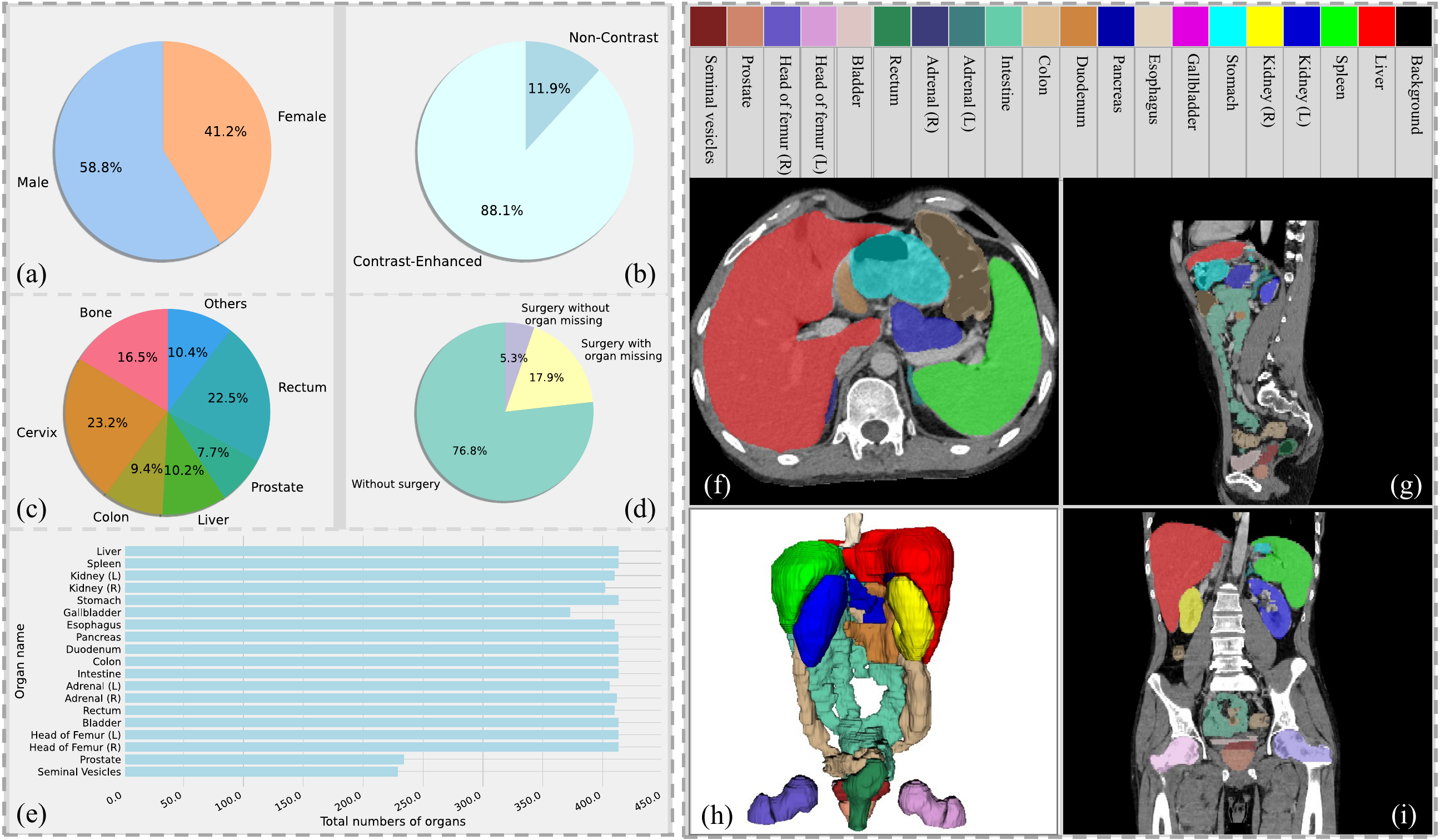}
    \caption{Data characteristics, distributions, and examples of the benchmark.}
    \label{fig:framework}
\end{figure*}
\section{RAOS}

\subsection{Dataset Collection}
In this study, we retrospectively collected 413 CT scans from 413 patients diagnosed with abdominal lesions from two institutions. All of them were produced by several different SIEMENS CT scanners with/without contrast agents to obey the clinical requirement. In addition, all 413 patients were treated with different strategies, such as surgery, adaptive radiation therapy, chemotherapy, or immunotherapy. It is worth pointing out that during the treatment processing each patient was scanned more than once, we selected the latest CT scan of each patient to construct the corner cases benchmark. The reason is that we found that some anatomical structures could change with the treatment like some organs were resected partially or whole after surgery and some lesions or organ sizes became small after radiotherapy, chemotherapy, or immunotherapy~\cite{rickmann2023halos}. 

\subsection{Dataset Statistical Analysis}
The detailed clinical characteristics of the RAOS dataset are presented in Fig.~\ref{fig:framework}. In the RAOS dataset, 58.8\% of patients (243) are male and the rest (170) are female. In addition, most of these CT scans' appearance is enhanced by injecting contrast agents before the scanning, which can enhance tumor or lesions in better view than non-contrast CT and further boost the clinical diagnosis. In this dataset, all patients were diagnosed with abdominal lesions where 23.2\%, 22.5\%, 16.5\%, 10.2\%, 9.4\%, and 7.7\% of them suffered from cervix, rectum, bone, liver, colon, and prostate, respectively and others include some metastatic tumors. Moreover, we further split the dataset into three subsets according to whether the treatment strategy has changed the anatomy: 1) without surgery (\textbf{\textit{SetA}}), patients treated by non-invasive methods without organ resection; 2) surgery without organ missing (\textbf{\textit{SetB}}), patients treated with surgery but just lesion and the neighbouring region was resected; 3) surgery with an organ missing (\textbf{\textit{SetC}}), patients treated with surgery and the organ with lesion was wholly resected. The total numbers of each organ were listed in Fig.~\ref{fig:framework}~(e), where the numbers of prostate and seminal vesicles are less than others as females do not have these two organs. Fig.~\ref{fig:framework}~(f-i) shows an example of a male patient in three different views.

\subsection{Ground Truth Generation}
To build a high-quality clinical challenging cases segmentation evaluation benchmark, all CT scans were manually delineated by a senior radiation oncologist ($>$ 8 years of clinical experience) using MIM Software~\cite{pukala2016benchmarking} under the latest RTOG delineation guideline~\cite{jabbour2014upper}. Then, another oncologist expert ($>$ 20 years of clinical experience) was invited to check these annotations and further discuss the region of unconsensus (especially for patients treated with resection surgery) to generate uniformity annotations. According to the above annotation procedure, all 413 CT scans (a total of 8000 3D organs) were comprehensively annotated with 17 (for female) /19 (for male) organs including the liver (Liv), spleen, left (L) and right (R) kidney (Kid), stomach (Sto), gallbladder (Gal), esophagus (Eso), pancreas (Pan), duodenum (Duo), colon (Col), intestine (Int), left and right adrenal (Adr), rectum (Rec), bladder (Bla), left and right head of the femur (Fem), prostate (Pro) and seminal vesicle (SVes). It is worth noticing that the RAOS dataset is annotated from scratch by senior oncologists, which is different from recent large-scale datasets using revised network predictions as ground truth~\cite{ma2021abdomenct,ji2022amos,wasserthal2022totalsegmentator}. So, we believe the RAOS can play an important role in model robustness and generalization evaluation on clinical corner cases and can also provide more organ types annotation for future research (such as prostate and seminal vesicle annotation in CT scans).

\section{Experiments and Results}
\subsection{Implementation Details}
\textbf{Data. }Considering the dataset consists of three subsets and the two surgery subsets have fewer patients, we used the \textbf{\textit{SetA}} as a development cohort for network training and internal evaluation containing 220 and 67 patients, respectively. Then, the \textbf{\textit{SetB}} and \textbf{\textit{SetC}} were used for robustness and generalization evaluation as they consisted of clinical corner cases. Afterwards, we also investigated the performance difference of baselines between the RAOS and several widely used abdominal organ segmentation datasets.
\\ \textbf{Metrics. }Two widely used metrics were used to measure the segmentation results in two different aspects~\cite{ma2021abdomenct}, 1) a voxel overlap-based metric Dice similarity coefficient (DSC); and 2) a boundary-based metric Normalized surface Dice (NSD) with a fixed tolerance distance of 2 $mm$. Moreover, a new metric about the organ hallucination ratio was introduced to measure false-positive predictions of resection organs for the surgery with organ missing subset~\cite{rickmann2023halos}. All of these metrics' implementations are publicly available.
\\ \textbf{Baselines. }To build a fair benchmark for clinical corner case segmentation evaluation, this work employed several state-of-the-art and publicly available segmentation methods as our baselines (nnUNet~\cite{isensee2021nnu}, 3DUXNET~\cite{lee20223d},REPUXNET~\cite{lee2023scaling},nnFormer~\cite{zhou2023nnformer},SwinUNETR~\cite{tang2022self},TransBTS~\cite{wang2021transbts}, UNETR~\cite{hatamizadeh2022unetr}). We ran all the experiments on a cluster with eight NVIDIA V100 GPUs and Pytorch 1.10~\cite{paszke2019pytorch} using these implementations' default hyper-parameters, pre-trained models and other experimental settings. 


\begin{table*}[t]
\centering
\normalsize
\scalebox{0.43}{\begin{tabular}{l|c|c|c|c|c|c|c|c|c|c|c|c|c|r}
\hline
\multirow{2}{*}{Method}& \multicolumn{2}{c}{nnUNet~\cite{isensee2021nnu}} & \multicolumn{2}{c}{REPUXNET~\cite{lee2023scaling}} & \multicolumn{2}{c}{3DUXNET~\cite{lee20223d}} & \multicolumn{2}{c}{nnFormer~\cite{zhou2023nnformer}} & \multicolumn{2}{c}{SwinUNETR~\cite{tang2022self}} & \multicolumn{2}{c}{TransBTS~\cite{wang2021transbts}} & \multicolumn{2}{c}{UNETR~\cite{hatamizadeh2022unetr}} \\
\cline{2-15} 
& \textit{DSC} (\%) &\textit{NSD (\%)}& \textit{DSC} (\%) &\textit{NSD (\%)}& \textit{DSC} (\%) &\textit{NSD (\%)}& \textit{DSC} (\%) &\textit{NSD (\%)}& \textit{DSC} (\%) &\textit{NSD (\%)}& \textit{DSC} (\%) &\textit{NSD (\%)}& \textit{DSC} (\%) &\textit{NSD (\%)} \\
\hline
Liv&96.65$\pm$0.98&90.59$\pm$3.12&96.51$\pm$1.03&90.04$\pm$3.45&96.43$\pm$1.14&89.65$\pm$3.74&95.95$\pm$1.23&87.43$\pm$4.39&96.4$\pm$1.21&89.69$\pm$3.56&96.03$\pm$1.26&87.48$\pm$3.86&95.95$\pm$1.2&87.11$\pm$3.78\\
Spl&95.93$\pm$1.09&96.91$\pm$2.23&95.85$\pm$1.0&96.73$\pm$2.31&95.58$\pm$1.55&96.17$\pm$3.04&94.69$\pm$1.51&93.93$\pm$4.41&95.72$\pm$0.97&96.44$\pm$1.85&95.23$\pm$1.35&95.31$\pm$3.03&94.67$\pm$1.52&93.09$\pm$4.5\\
LKid&95.8$\pm$1.18&96.44$\pm$2.18&95.59$\pm$1.22&95.13$\pm$3.63&95.51$\pm$1.42&95.6$\pm$3.03&94.57$\pm$1.61&93.62$\pm$3.43&95.56$\pm$1.11&95.77$\pm$2.21&95.18$\pm$1.22&95.08$\pm$2.71&94.73$\pm$1.45&93.47$\pm$3.17\\
RKid&95.56$\pm$1.36&96.13$\pm$2.53&95.21$\pm$1.9&94.61$\pm$5.51&95.52$\pm$1.37&96.0$\pm$2.72&94.53$\pm$1.78&93.56$\pm$4.0&95.35$\pm$1.32&95.49$\pm$2.63&94.87$\pm$1.31&93.77$\pm$3.09&94.62$\pm$2.05&93.52$\pm$4.4\\
Sto&92.77$\pm$3.69&86.99$\pm$5.69&91.92$\pm$4.17&84.09$\pm$6.83&92.17$\pm$3.53&84.58$\pm$7.08&90.78$\pm$3.59&79.73$\pm$7.18&92.04$\pm$3.56&83.97$\pm$6.6&91.71$\pm$3.37&82.71$\pm$6.17&90.21$\pm$4.46&78.77$\pm$7.9\\
Gal&80.63$\pm$17.36&83.87$\pm$17.86&76.02$\pm$16.78&75.04$\pm$18.08&77.15$\pm$18.65&78.78$\pm$19.65&71.07$\pm$17.35&69.35$\pm$18.13&75.76$\pm$18.25&76.12$\pm$20.32&72.24$\pm$20.46&72.13$\pm$20.81&68.11$\pm$20.5&67.03$\pm$20.84\\
Eso&83.23$\pm$5.51&90.78$\pm$5.49&78.96$\pm$8.42&85.7$\pm$8.07&78.57$\pm$9.45&86.08$\pm$8.76&71.6$\pm$11.06&79.24$\pm$11.22&78.18$\pm$8.84&85.89$\pm$8.35&75.53$\pm$8.23&83.16$\pm$7.68&70.69$\pm$13.5&77.31$\pm$13.32\\
Pan&83.28$\pm$8.2&82.89$\pm$8.8&81.82$\pm$8.77&80.75$\pm$9.33&81.62$\pm$8.92&80.65$\pm$9.56&77.16$\pm$9.1&73.09$\pm$9.47&81.74$\pm$8.45&80.39$\pm$9.58&78.8$\pm$9.23&75.55$\pm$9.84&77.88$\pm$8.64&73.43$\pm$9.29\\
Duo&72.05$\pm$16.02&73.0$\pm$14.71&68.25$\pm$15.21&68.57$\pm$14.16&68.42$\pm$15.22&68.38$\pm$13.68&64.12$\pm$14.67&61.69$\pm$12.55&68.72$\pm$15.45&68.75$\pm$14.43&65.31$\pm$15.37&65.0$\pm$13.66&63.26$\pm$13.95&59.14$\pm$13.04\\
Col&87.1$\pm$8.39&84.74$\pm$10.09&85.4$\pm$8.26&80.79$\pm$9.91&85.25$\pm$8.36&80.22$\pm$10.14&81.65$\pm$8.39&72.27$\pm$10.14&84.96$\pm$8.12&79.3$\pm$9.85&84.12$\pm$8.19&77.45$\pm$10.13&80.72$\pm$7.31&69.74$\pm$9.1\\
Int&87.94$\pm$6.54&88.21$\pm$7.3&86.6$\pm$6.95&85.6$\pm$7.77&86.19$\pm$7.43&85.07$\pm$8.47&84.05$\pm$5.56&80.33$\pm$8.24&86.39$\pm$6.28&84.57$\pm$7.73&85.51$\pm$5.99&83.15$\pm$7.89&83.4$\pm$5.9&78.84$\pm$8.34\\
RAdr&71.82$\pm$15.22&85.04$\pm$16.7&69.63$\pm$15.28&82.83$\pm$16.5&70.58$\pm$14.79&84.29$\pm$15.34&59.2$\pm$16.48&73.56$\pm$17.66&69.1$\pm$14.99&82.73$\pm$16.47&62.84$\pm$14.05&75.85$\pm$15.39&63.14$\pm$14.94&77.95$\pm$16.27\\
LAdr&72.68$\pm$18.84&84.15$\pm$20.15&69.93$\pm$18.41&81.9$\pm$19.4&70.81$\pm$19.01&82.96$\pm$19.91&55.99$\pm$20.63&69.29$\pm$21.57&69.63$\pm$18.09&80.76$\pm$19.31&64.87$\pm$17.93&76.48$\pm$19.6&64.66$\pm$17.52&76.74$\pm$18.62\\
Rec&83.1$\pm$11.16&79.59$\pm$12.1&81.53$\pm$11.97&75.98$\pm$12.9&81.56$\pm$12.24&76.63$\pm$13.21&71.93$\pm$12.12&59.73$\pm$11.54&80.96$\pm$10.66&74.85$\pm$11.87&76.92$\pm$10.35&66.83$\pm$11.68&76.05$\pm$11.04&64.92$\pm$12.67\\
Bla&94.86$\pm$3.81&90.88$\pm$6.76&94.21$\pm$4.33&89.33$\pm$7.37&94.08$\pm$4.99&89.01$\pm$7.65&91.49$\pm$9.56&82.71$\pm$11.55&93.73$\pm$5.99&88.18$\pm$8.44&92.7$\pm$6.9&85.0$\pm$9.31&92.12$\pm$6.18&82.03$\pm$8.29\\
LFem&84.99$\pm$17.07&81.24$\pm$16.43&92.27$\pm$4.22&89.47$\pm$5.32&92.18$\pm$4.4&89.61$\pm$5.37&89.95$\pm$3.95&83.36$\pm$5.8&91.79$\pm$3.89&88.34$\pm$4.77&91.62$\pm$4.03&87.59$\pm$5.05&91.0$\pm$3.55&85.55$\pm$4.67\\
RFem&86.79$\pm$9.11&81.69$\pm$11.24&91.68$\pm$4.32&88.5$\pm$5.22&91.66$\pm$3.89&88.3$\pm$4.46&89.18$\pm$3.48&81.96$\pm$5.43&91.74$\pm$3.98&88.31$\pm$4.89&91.32$\pm$4.01&86.93$\pm$5.08&91.1$\pm$3.94&86.4$\pm$5.12\\
Pro&93.01$\pm$6.94&88.96$\pm$11.62&60.22$\pm$41.11&53.75$\pm$37.96&81.33$\pm$28.78&75.76$\pm$28.93&84.75$\pm$17.81&75.78$\pm$24.1&68.73$\pm$37.77&62.06$\pm$36.01&86.65$\pm$16.33&77.48$\pm$23.44&71.57$\pm$34.67&63.19$\pm$33.84\\
SVes&88.23$\pm$14.58&88.5$\pm$14.84&62.17$\pm$37.62&61.42$\pm$37.67&67.03$\pm$35.91&66.6$\pm$36.21&77.48$\pm$23.38&75.55$\pm$24.96&63.63$\pm$37.04&63.91$\pm$37.3&79.98$\pm$22.25&79.26$\pm$22.84&55.18$\pm$36.99&54.02$\pm$36.82\\
\hline
Mean $\uparrow$&86.65&86.87&82.83&82.12&84.3&83.91&81.06&78.22&83.17&82.4&83.23&81.38&79.95&76.96\\
\hline
\end{tabular}}
\caption{Comparison between SOTA methods in terms of $DSC (\%)$ and $NSD (\%)$ on the \textbf{\textit{SetA}}.}
\label{tab:witout-surgery}
\end{table*}

\begin{table*}[t]
\centering
\normalsize
\scalebox{0.43}{\begin{tabular}{l|c|c|c|c|c|c|c|c|c|c|c|c|c|r}
\hline
\multirow{2}{*}{Method}& \multicolumn{2}{c}{nnUNet~\cite{isensee2021nnu}} & \multicolumn{2}{c}{REPUXNET~\cite{lee2023scaling}} & \multicolumn{2}{c}{3DUXNET~\cite{lee20223d}} & \multicolumn{2}{c}{nnFormer~\cite{zhou2023nnformer}} & \multicolumn{2}{c}{SwinUNETR~\cite{tang2022self}} & \multicolumn{2}{c}{TransBTS~\cite{wang2021transbts}} & \multicolumn{2}{c}{UNETR~\cite{hatamizadeh2022unetr}} \\
\cline{2-15} 
& \textit{DSC} (\%) &\textit{NSD (\%)}& \textit{DSC} (\%) &\textit{NSD (\%)}& \textit{DSC} (\%) &\textit{NSD (\%)}& \textit{DSC} (\%) &\textit{NSD (\%)}& \textit{DSC} (\%) &\textit{NSD (\%)}& \textit{DSC} (\%) &\textit{NSD (\%)}& \textit{DSC} (\%) &\textit{NSD (\%)} \\
\hline
Liv&96.06$\pm$2.57&88.01$\pm$8.73&96.34$\pm$1.0&88.34$\pm$5.83&96.4$\pm$0.84&88.51$\pm$5.52&95.75$\pm$1.65&86.32$\pm$6.7&96.36$\pm$1.08&88.62$\pm$5.55&95.77$\pm$1.47&85.26$\pm$8.72&95.93$\pm$0.88&85.92$\pm$5.48\\
Spl&95.32$\pm$2.41&95.32$\pm$6.25&95.41$\pm$1.27&94.61$\pm$5.29&95.38$\pm$1.28&94.44$\pm$4.88&94.64$\pm$1.44&92.94$\pm$5.18&95.28$\pm$1.85&94.64$\pm$6.68&95.35$\pm$1.3&95.09$\pm$3.56&94.61$\pm$1.46&91.95$\pm$5.4\\
LKid&94.46$\pm$5.04&95.13$\pm$5.72&95.14$\pm$1.39&94.23$\pm$4.06&95.31$\pm$1.16&95.44$\pm$2.33&93.93$\pm$1.66&91.85$\pm$4.46&95.33$\pm$1.22&95.29$\pm$2.26&94.73$\pm$1.3&94.34$\pm$2.82&94.5$\pm$1.25&92.93$\pm$2.84\\
RKid&94.78$\pm$4.25&95.38$\pm$5.28&95.27$\pm$1.54&94.54$\pm$4.49&95.53$\pm$1.18&96.16$\pm$2.1&94.48$\pm$1.6&94.01$\pm$3.25&95.39$\pm$0.98&95.8$\pm$1.77&94.66$\pm$1.25&93.51$\pm$2.63&94.94$\pm$1.06&94.54$\pm$2.48\\
Sto&91.94$\pm$4.46&82.98$\pm$11.58&92.0$\pm$3.54&81.59$\pm$9.57&90.68$\pm$6.29&80.92$\pm$11.48&89.35$\pm$6.77&75.66$\pm$13.35&90.1$\pm$8.43&79.79$\pm$14.7&91.08$\pm$4.53&79.94$\pm$10.92&89.02$\pm$7.66&75.26$\pm$13.79\\
Gal&78.03$\pm$14.33&81.28$\pm$15.22&71.82$\pm$14.84&70.38$\pm$17.15&76.25$\pm$12.82&77.76$\pm$16.48&60.22$\pm$25.56&60.53$\pm$26.61&74.6$\pm$14.6&76.57$\pm$18.91&67.77$\pm$18.24&67.63$\pm$19.25&65.56$\pm$18.42&64.66$\pm$20.2\\
Eso&81.59$\pm$4.94&88.06$\pm$6.18&78.36$\pm$5.74&83.76$\pm$7.42&78.49$\pm$6.6&84.35$\pm$8.14&71.46$\pm$9.8&78.39$\pm$10.44&77.65$\pm$7.86&83.88$\pm$9.12&74.92$\pm$8.08&81.28$\pm$10.14&71.96$\pm$10.96&76.88$\pm$13.32\\
Pan&84.11$\pm$6.65&82.57$\pm$8.62&83.4$\pm$6.79&81.29$\pm$8.78&83.05$\pm$7.1&81.16$\pm$8.98&77.08$\pm$9.04&71.12$\pm$10.79&81.87$\pm$8.46&79.14$\pm$10.33&79.6$\pm$7.19&74.62$\pm$8.91&79.27$\pm$7.82&73.9$\pm$9.33\\
Duo&70.63$\pm$11.85&71.61$\pm$12.7&70.56$\pm$11.59&70.05$\pm$12.62&68.97$\pm$13.28&69.21$\pm$13.48&65.33$\pm$10.83&62.35$\pm$12.46&69.08$\pm$12.88&68.49$\pm$13.28&67.03$\pm$14.5&66.88$\pm$14.88&66.94$\pm$10.56&61.63$\pm$10.92\\
Col&84.06$\pm$10.1&80.02$\pm$13.33&82.08$\pm$8.03&75.52$\pm$11.85&81.61$\pm$8.66&75.76$\pm$12.27&76.98$\pm$10.06&66.61$\pm$11.61&80.98$\pm$9.98&74.31$\pm$13.33&81.02$\pm$8.66&73.7$\pm$11.42&76.7$\pm$8.42&64.74$\pm$12.76\\
Int&86.55$\pm$6.45&86.05$\pm$10.29&84.72$\pm$7.02&83.31$\pm$10.99&84.57$\pm$7.55&83.06$\pm$11.71&81.86$\pm$6.68&78.34$\pm$10.62&84.37$\pm$7.48&82.52$\pm$11.56&83.91$\pm$6.88&81.86$\pm$10.52&81.8$\pm$6.63&77.92$\pm$10.62\\
RAdr&71.0$\pm$14.42&85.22$\pm$15.53&70.12$\pm$14.36&83.5$\pm$15.82&69.22$\pm$15.06&83.18$\pm$16.59&60.56$\pm$15.98&74.85$\pm$18.79&68.53$\pm$13.18&82.1$\pm$14.16&61.65$\pm$13.24&75.19$\pm$14.54&63.97$\pm$15.09&78.66$\pm$15.52\\
LAdr&75.17$\pm$9.2&86.66$\pm$9.93&72.86$\pm$10.1&85.24$\pm$10.26&73.49$\pm$9.83&85.55$\pm$9.97&57.28$\pm$17.31&70.18$\pm$18.68&71.7$\pm$9.77&82.87$\pm$10.41&66.88$\pm$10.18&78.9$\pm$11.31&65.7$\pm$11.28&76.94$\pm$12.45\\
Rec&80.25$\pm$9.49&76.07$\pm$12.56&78.9$\pm$9.22&73.29$\pm$11.0&78.54$\pm$10.59&73.56$\pm$13.55&73.95$\pm$10.35&65.16$\pm$11.17&77.53$\pm$10.62&72.48$\pm$13.26&74.82$\pm$10.94&67.43$\pm$13.45&73.52$\pm$9.77&64.55$\pm$12.95\\
Bla&93.12$\pm$10.75&89.11$\pm$11.33&92.83$\pm$11.13&88.76$\pm$10.8&92.77$\pm$11.28&88.28$\pm$11.83&90.87$\pm$12.83&83.06$\pm$12.51&92.68$\pm$10.85&87.76$\pm$11.58&92.34$\pm$10.33&86.43$\pm$10.85&91.53$\pm$11.33&82.81$\pm$13.49\\
LFem&70.52$\pm$32.47&68.31$\pm$30.1&92.2$\pm$4.9&89.44$\pm$6.62&92.21$\pm$5.21&89.29$\pm$7.76&89.49$\pm$4.3&81.97$\pm$7.7&91.44$\pm$4.35&87.2$\pm$6.59&91.47$\pm$4.64&87.26$\pm$6.95&90.65$\pm$3.82&84.56$\pm$6.27\\
RFem&83.35$\pm$10.91&78.13$\pm$12.43&91.64$\pm$4.15&88.32$\pm$5.56&91.63$\pm$3.87&87.81$\pm$4.52&88.91$\pm$3.36&80.94$\pm$5.95&91.85$\pm$4.15&88.76$\pm$5.63&91.34$\pm$4.06&87.12$\pm$5.53&90.88$\pm$3.91&85.9$\pm$5.08\\
Pro&87.31$\pm$10.16&79.7$\pm$15.26&67.48$\pm$32.98&57.88$\pm$30.62&85.76$\pm$11.68&77.08$\pm$18.09&76.48$\pm$22.09&64.85$\pm$25.68&75.28$\pm$26.42&64.19$\pm$27.69&78.6$\pm$21.84&68.0$\pm$24.21&76.54$\pm$22.56&64.29$\pm$25.49\\
SVes&79.99$\pm$23.38&81.0$\pm$22.75&66.71$\pm$31.67&66.2$\pm$31.67&72.06$\pm$28.32&71.92$\pm$28.49&57.79$\pm$35.3&59.85$\pm$32.96&66.56$\pm$33.23&67.3$\pm$32.89&67.1$\pm$32.37&67.7$\pm$30.69&56.04$\pm$34.38&55.98$\pm$33.96\\
\hline
Mean$\uparrow$&84.12&83.72&83.04&81.59&84.31&83.34&78.76&75.74&82.98&81.67&81.58&79.59&80.0&76.53\\
\hline
\end{tabular}}
\caption{Comparison between SOTA methods in terms of $DSC (\%)$ and $NSD (\%)$ on the \textbf{\textit{SetB}}.}
\label{tab:surgery-without-missing}
\end{table*}

\begin{table*}[t]
\centering
\normalsize
\scalebox{0.43}{\begin{tabular}{l|c|c|c|c|c|c|c|c|c|c|c|c|c|r}
\hline
\multirow{2}{*}{Method}& \multicolumn{2}{c}{nnUNet~\cite{isensee2021nnu}} & \multicolumn{2}{c}{REPUXNET~\cite{lee2023scaling}} & \multicolumn{2}{c}{3DUXNET~\cite{lee20223d}} & \multicolumn{2}{c}{nnFormer~\cite{zhou2023nnformer}} & \multicolumn{2}{c}{SwinUNETR~\cite{tang2022self}} & \multicolumn{2}{c}{TransBTS~\cite{wang2021transbts}} & \multicolumn{2}{c}{UNETR~\cite{hatamizadeh2022unetr}} \\
\cline{2-15} 
& \textit{DSC} (\%) &\textit{NSD (\%)}& \textit{DSC} (\%) &\textit{NSD (\%)}& \textit{DSC} (\%) &\textit{NSD (\%)}& \textit{DSC} (\%) &\textit{NSD (\%)}& \textit{DSC} (\%) &\textit{NSD (\%)}& \textit{DSC} (\%) &\textit{NSD (\%)}& \textit{DSC} (\%) &\textit{NSD (\%)} \\
\hline
Liv&95.35$\pm$8.73&88.21$\pm$9.78&94.88$\pm$9.77&86.74$\pm$10.97&94.68$\pm$10.59&86.25$\pm$11.31&94.69$\pm$8.72&85.0$\pm$9.91&94.82$\pm$9.75&86.35$\pm$10.78&94.84$\pm$8.83&85.68$\pm$9.2&94.69$\pm$7.57&84.29$\pm$10.02\\
Spl&95.37$\pm$2.08&96.16$\pm$4.17&94.6$\pm$4.1&94.42$\pm$7.26&94.06$\pm$4.26&92.33$\pm$9.2&93.67$\pm$3.16&92.13$\pm$6.57&94.78$\pm$2.34&94.24$\pm$5.78&94.4$\pm$2.35&93.44$\pm$5.66&93.73$\pm$3.26&91.19$\pm$7.01\\
LKid&90.95$\pm$18.92&90.91$\pm$19.31&90.65$\pm$18.87&89.09$\pm$19.41&90.09$\pm$19.12&88.4$\pm$20.23&90.91$\pm$15.83&88.88$\pm$16.65&90.24$\pm$19.04&88.81$\pm$19.71&90.53$\pm$18.71&89.19$\pm$18.9&89.1$\pm$19.04&86.25$\pm$19.85\\
RKid&89.36$\pm$21.98&89.31$\pm$22.5&78.17$\pm$35.35&77.2$\pm$35.45&82.02$\pm$32.03&81.72$\pm$32.25&89.16$\pm$20.25&87.71$\pm$21.13&79.28$\pm$34.01&78.77$\pm$34.25&79.08$\pm$33.69&77.0$\pm$33.86&77.8$\pm$34.59&75.91$\pm$34.45\\
Sto&90.75$\pm$7.93&83.16$\pm$10.58&89.07$\pm$8.37&78.51$\pm$13.61&87.96$\pm$10.16&77.43$\pm$14.23&88.25$\pm$8.86&75.38$\pm$12.54&88.63$\pm$9.31&77.7$\pm$12.89&89.76$\pm$7.27&78.56$\pm$10.68&87.04$\pm$8.89&72.54$\pm$13.67\\
Gal&54.69$\pm$42.8&55.7$\pm$43.43&33.81$\pm$38.65&33.11$\pm$37.57&40.2$\pm$41.14&40.44$\pm$41.63&42.32$\pm$40.04&41.08$\pm$39.28&40.19$\pm$40.99&40.17$\pm$40.65&33.92$\pm$38.87&34.17$\pm$38.65&29.8$\pm$35.52&28.95$\pm$33.86\\
Eso&80.51$\pm$9.84&88.11$\pm$9.88&73.6$\pm$15.88&79.7$\pm$16.75&73.08$\pm$16.81&80.47$\pm$17.29&68.98$\pm$15.81&75.86$\pm$16.61&73.48$\pm$17.13&80.72$\pm$17.65&70.65$\pm$16.13&77.87$\pm$17.15&67.08$\pm$18.84&73.21$\pm$19.86\\
Pan&81.05$\pm$12.36&79.52$\pm$13.05&77.66$\pm$13.28&74.63$\pm$14.67&76.39$\pm$14.01&73.45$\pm$15.19&72.47$\pm$14.84&67.8$\pm$14.49&77.32$\pm$13.66&74.15$\pm$15.06&75.26$\pm$13.09&70.74$\pm$14.44&71.87$\pm$14.31&66.33$\pm$14.45\\
Duo&64.79$\pm$17.55&64.32$\pm$17.55&59.43$\pm$18.65&58.32$\pm$17.85&58.49$\pm$17.42&57.36$\pm$17.21&57.8$\pm$16.81&54.95$\pm$16.02&58.16$\pm$18.68&57.82$\pm$18.16&59.14$\pm$18.93&56.4$\pm$18.3&55.92$\pm$16.32&51.25$\pm$15.73\\
Col&84.9$\pm$10.69&80.77$\pm$13.04&81.56$\pm$9.79&74.61$\pm$13.02&81.15$\pm$10.23&74.07$\pm$12.78&80.08$\pm$8.64&69.63$\pm$11.13&80.25$\pm$9.67&72.22$\pm$12.39&81.8$\pm$8.51&73.6$\pm$11.76&76.36$\pm$8.76&63.92$\pm$11.37\\
Int&86.07$\pm$7.65&84.07$\pm$10.31&83.62$\pm$7.93&79.59$\pm$11.83&83.08$\pm$8.45&78.82$\pm$12.06&81.63$\pm$8.34&75.89$\pm$11.75&82.81$\pm$8.27&78.21$\pm$12.04&83.24$\pm$7.03&78.42$\pm$11.31&80.15$\pm$8.14&72.88$\pm$12.43\\
RAdr&65.65$\pm$25.07&76.92$\pm$27.81&59.94$\pm$26.39&71.07$\pm$29.37&60.44$\pm$26.88&72.37$\pm$29.47&53.73$\pm$26.0&65.95$\pm$29.35&59.85$\pm$26.64&71.85$\pm$28.93&52.98$\pm$25.37&64.52$\pm$29.15&54.38$\pm$25.76&67.02$\pm$28.47\\
LAdr&72.46$\pm$15.78&83.46$\pm$16.52&66.04$\pm$19.35&77.72$\pm$19.96&68.06$\pm$17.9&79.51$\pm$19.29&51.16$\pm$26.37&62.06$\pm$30.4&66.8$\pm$15.86&77.68$\pm$16.49&60.31$\pm$18.51&70.72$\pm$20.57&60.75$\pm$18.57&71.77$\pm$19.7\\
Rec&77.26$\pm$21.4&73.87$\pm$21.36&74.63$\pm$23.52&70.05$\pm$23.24&74.14$\pm$23.14&69.02$\pm$23.3&65.32$\pm$22.85&55.39$\pm$20.52&73.02$\pm$22.85&67.86$\pm$22.55&71.23$\pm$22.83&63.28$\pm$21.9&68.27$\pm$22.44&59.17$\pm$21.47\\
Bla&92.57$\pm$9.36&87.12$\pm$13.07&91.67$\pm$8.99&84.35$\pm$13.65&91.14$\pm$10.47&83.35$\pm$15.44&89.01$\pm$12.38&78.8$\pm$16.66&91.42$\pm$10.14&83.87$\pm$14.49&90.51$\pm$9.95&80.96$\pm$14.86&88.42$\pm$13.02&76.14$\pm$16.99\\
LFem&79.64$\pm$22.58&76.56$\pm$21.06&90.14$\pm$11.83&87.0$\pm$12.1&90.37$\pm$11.41&86.75$\pm$12.26&87.94$\pm$11.66&80.58$\pm$11.72&90.26$\pm$11.23&86.2$\pm$11.46&89.64$\pm$11.42&85.02$\pm$11.98&89.27$\pm$11.18&83.39$\pm$11.36\\
RFem&80.67$\pm$21.41&76.47$\pm$19.87&89.55$\pm$12.87&86.01$\pm$13.26&89.44$\pm$12.81&85.48$\pm$13.01&86.77$\pm$12.71&78.98$\pm$13.31&89.4$\pm$13.38&85.52$\pm$14.01&89.42$\pm$11.61&84.3$\pm$12.26&89.07$\pm$12.13&83.83$\pm$12.94\\
Pro&80.23$\pm$31.1&75.25$\pm$31.64&45.94$\pm$41.89&39.09$\pm$37.18&68.18$\pm$37.7&61.27$\pm$37.33&75.99$\pm$30.08&67.43$\pm$33.03&53.65$\pm$41.89&46.83$\pm$38.5&70.1$\pm$34.91&61.22$\pm$36.62&55.05$\pm$40.62&46.44$\pm$37.33\\
SVes&68.13$\pm$38.82&68.3$\pm$38.87&43.79$\pm$41.14&42.61$\pm$40.65&46.26$\pm$41.7&45.35$\pm$41.33&69.11$\pm$35.28&67.67$\pm$35.5&39.25$\pm$39.99&38.54$\pm$39.6&52.79$\pm$41.14&52.47$\pm$41.06&37.46$\pm$38.96&36.83$\pm$38.06\\
\hline
Mean $\uparrow$&80.55&79.9&74.67&72.83&76.27&74.41&75.74&72.17&74.93&73.03&75.24&72.5&71.91&67.96\\
\hline
\end{tabular}}
\caption{Comparison between SOTA methods in terms of $DSC (\%)$ and $NSD (\%)$ on the \textbf{\textit{SetC}}.}
\label{tab:surgery-with-missing}
\end{table*}

\begin{table}[t]
\centering
\normalsize
\scalebox{0.55}{\begin{tabular}{l|c|c|c|c|c|c|r}
\hline
{Method}&{nnUNet~\cite{isensee2021nnu}}&{REPUXNET~\cite{lee2023scaling}}& {3DUXNET~\cite{lee20223d}}&{nnFormer~\cite{zhou2023nnformer}}&{SwinUNETR~\cite{tang2022self}}&{TransBTS~\cite{wang2021transbts}}& {UNETR~\cite{hatamizadeh2022unetr}}\\
\hline
LKid&1.0&1.0&1.0&0.67&1.0&1.0&1.0\\
RKid&0.33&1.0&0.75&0.18&0.92&0.91&1.0\\
Gal&0.68&1.0&0.93&0.78&0.85&0.97&1.0\\
RAdr&0.75&0.88&0.75& 0.78&0.75&1.0&0.78\\
LAdr&1.0&1.0&1.0&1.0&1.0&1.0&1.0\\
Rec&0.67&1.0&1.0&1.0&1.0&1.0&1.0\\
Pro&0.24&0.89&0.43&0.18&0.73&0.32&0.65\\
SVes&0.38&0.77&0.72&0.23&0.84&0.52&0.79\\
\hline
Mean $\downarrow$&0.63&0.94&0.82&0.6&0.89&0.84&0.9\\
\hline
\end{tabular}}
\caption{The organ hallucinations ratio of seven different methods on missing organ cases.}
\label{tab:hall_ratio}
\end{table}

\begin{table}[t]
\centering
\normalsize
\scalebox{0.5}{\begin{tabular}{l|c|c|c|c|c|c|c|r}
\hline
{Dataset}& \multicolumn{4}{c}{RAOS $\rightarrow$ BTCV~\cite{landman2015miccai}}& \multicolumn{4}{c}{RAOS $\rightarrow$ AbdomenCT-1K~\cite{ma2021abdomenct}} \\
\hline
\multirow{2}{*}{Method}& \multicolumn{2}{c}{nnUNet~\cite{isensee2021nnu}} & \multicolumn{2}{c}{3DUXNET~\cite{lee20223d}} & \multicolumn{2}{c}{nnUNet~\cite{isensee2021nnu}} & \multicolumn{2}{c}{3DUXNET~\cite{lee20223d}} \\
\cline{2-9} 
& \textit{DSC} (\%) &\textit{NSD (\%)}& \textit{DSC} (\%) &\textit{NSD (\%)}& \textit{DSC} (\%) &\textit{NSD (\%)}& \textit{DSC} (\%) &\textit{NSD (\%)}\\
\hline
Liv&93.85$\pm$6.46&81.03$\pm$9.02&94.02$\pm$3.62&80.06$\pm$9.72&95.83$\pm$0.71&83.66$\pm$4.57&95.08$\pm$1.05&81.48$\pm$4.17\\
Spl&86.82$\pm$16.13&82.72$\pm$17.44&85.95$\pm$14.83&77.94$\pm$18.99&94.73$\pm$4.25&93.45$\pm$5.61&92.49$\pm$7.4&87.32$\pm$10.85\\
LKid&88.61$\pm$9.34&85.42$\pm$9.8&87.14$\pm$14.5&83.12$\pm$14.03&92.18$\pm$3.86&86.28$\pm$6.01&91.7$\pm$9.44&86.14$\pm$8.5\\
RKid&87.21$\pm$16.46&84.44$\pm$14.94&87.57$\pm$15.49&82.15$\pm$13.28&92.17$\pm$6.59&87.3$\pm$6.95&91.49$\pm$7.51&85.33$\pm$8.36\\
Sto&82.58$\pm$17.46&69.75$\pm$18.05&74.45$\pm$21.58&61.25$\pm$20.43&89.51$\pm$9.55&80.21$\pm$11.82&90.25$\pm$5.84&79.97$\pm$8.76\\
Gal&68.41$\pm$30.91&68.27$\pm$30.42&62.73$\pm$32.35&63.17$\pm$30.44&87.21$\pm$5.18&87.59$\pm$9.03&82.44$\pm$12.16&81.64$\pm$14.29\\
Eso&78.64$\pm$7.37&84.41$\pm$8.19&74.86$\pm$11.8&80.38$\pm$12.83&79.49$\pm$7.54&82.34$\pm$7.7&67.19$\pm$18.49&70.48$\pm$17.6\\
Pan&77.16$\pm$10.29&72.84$\pm$12.95&61.48$\pm$23.45&58.51$\pm$22.75&79.79$\pm$7.15&69.08$\pm$11.93&73.24$\pm$13.72&62.06$\pm$15.56\\
RAdr&69.38$\pm$6.32&86.75$\pm$7.28&66.01$\pm$8.37&83.96$\pm$9.59&77.67$\pm$7.5&92.48$\pm$7.39&70.07$\pm$16.37&86.27$\pm$16.58\\
LAdr&68.57$\pm$14.39&85.02$\pm$15.5&64.78$\pm$16.22&79.92$\pm$18.73&75.09$\pm$5.81&92.07$\pm$6.32&74.13$\pm$5.94&88.86$\pm$7.46\\
\hline
Mean $\uparrow$&80.12&80.06&75.9&75.05&86.37&85.45&82.81&80.95\\
\hline
\end{tabular}}
\caption{The generalizable results from RAOS to BTCV and AbdomenCT-1K.}
\label{tab:raos2others}
\end{table}

\begin{table}[t]
\centering
\normalsize
\scalebox{0.55}{\begin{tabular}{l|c|c|c|c|c|c|c|r}
\hline
\multirow{2}{*}{Dataset}& \multicolumn{2}{c}{RAOS $\rightarrow$ AMOS~\cite{ji2022amos}} & \multicolumn{2}{c}{AMOS~\cite{ji2022amos} $\rightarrow$ RAOS (\textbf{\textit{SetA}})} & \multicolumn{2}{c}{AMOS~\cite{ji2022amos} $\rightarrow$ RAOS (\textbf{\textit{SetB}})} & \multicolumn{2}{c}{AMOS~\cite{ji2022amos} $\rightarrow$ RAOS (\textbf{\textit{SetC}})} \\
\cline{2-9} 
& \textit{DSC} (\%) &\textit{NSD (\%)}& \textit{DSC} (\%) &\textit{NSD (\%)}& \textit{DSC} (\%) &\textit{NSD (\%)}& \textit{DSC} (\%) &\textit{NSD (\%)}\\
\hline
Liv&95.72$\pm$2.53&88.31$\pm$7.84&96.12$\pm$1.08&88.75$\pm$3.72&95.15$\pm$4.41&87.49$\pm$6.31&94.74$\pm$9.19&86.3$\pm$9.45\\
Spl&94.24$\pm$3.47&93.31$\pm$6.95&94.97$\pm$1.19&95.37$\pm$2.65&92.56$\pm$9.43&93.07$\pm$9.46&94.3$\pm$2.28&93.91$\pm$4.91\\
LKid&92.58$\pm$3.33&91.7$\pm$4.75&93.36$\pm$2.02&92.44$\pm$3.63&91.71$\pm$5.7&90.67$\pm$6.98&90.24$\pm$15.21&88.12$\pm$15.52\\
RKid&92.79$\pm$2.98&91.84$\pm$4.74&93.38$\pm$2.21&92.62$\pm$4.43&92.52$\pm$4.41&91.76$\pm$6.29&89.95$\pm$16.56&88.49$\pm$17.36\\
Sto&87.61$\pm$13.65&81.03$\pm$16.83&91.64$\pm$3.16&82.96$\pm$6.79&91.42$\pm$2.85&80.49$\pm$8.43&90.06$\pm$5.51&79.45$\pm$9.72\\
Gal&74.6$\pm$28.12&77.89$\pm$28.93&74.58$\pm$23.18&77.85$\pm$23.23&72.39$\pm$21.74&76.71$\pm$20.42&50.49$\pm$42.8&51.85$\pm$43.92\\
Eso&79.28$\pm$14.31&88.36$\pm$14.69&79.32$\pm$6.19&87.35$\pm$6.52&77.85$\pm$6.4&85.09$\pm$7.78&75.68$\pm$13.11&83.67$\pm$13.82\\
Pan&81.77$\pm$9.23&82.0$\pm$10.74&80.39$\pm$8.2&79.21$\pm$9.47&82.03$\pm$5.51&79.89$\pm$7.57&77.81$\pm$12.5&75.34$\pm$14.06\\
Duo&69.88$\pm$13.43&70.83$\pm$14.53&67.42$\pm$17.21&69.83$\pm$16.59&66.75$\pm$16.74&70.53$\pm$16.5&61.09$\pm$18.2&61.81$\pm$18.73\\
RAdr&70.12$\pm$8.39&88.69$\pm$8.09&65.87$\pm$14.82&82.39$\pm$17.11&64.29$\pm$13.13&80.96$\pm$14.93&59.29$\pm$25.27&72.84$\pm$30.15\\
LAdr&68.52$\pm$9.06&86.93$\pm$9.89&64.43$\pm$18.03&79.25$\pm$20.67&68.33$\pm$8.38&83.49$\pm$10.15&65.74$\pm$15.56&80.03$\pm$17.49\\
Bla&78.51$\pm$23.58&74.94$\pm$23.13&92.37$\pm$10.06&86.21$\pm$11.29&93.36$\pm$4.54&86.8$\pm$7.75&90.75$\pm$13.23&83.14$\pm$14.9\\
\hline
Mean $\uparrow$&82.14&84.65&82.82&84.52&82.36&83.91&78.34&78.74\\
\hline
\end{tabular}}
\caption{The cross-evaluation results between the RAOS and AMOS based on the nnUNet.}
\label{tab:cross}
\end{table}

\subsection{Results}
\textbf{Results on the \textbf{\textit{SetA}}. }We first investigated seven state-of-the-art 3D medical image segmentation methods on the \textbf{\textit{SetA}}, including a CNN-based nnUNet~\cite{isensee2021nnu} and six transformer-based REPUXNET~\cite{lee2023scaling}, 3DUXNET~\cite{lee20223d}, nnFormer~\cite{zhou2023nnformer}, SwinUNETR~\cite{tang2022self}, TransBTS~\cite{wang2021transbts} and UNETR~\cite{hatamizadeh2022unetr}. Table~\ref{tab:witout-surgery} presented the quantitative results in terms of DSC and NSD showing that nnUNet was still the best solution for abdominal organ segmentation which outperformed all transformer-based methods. Besides, we found that the nnUNet predictions of several large-size organs in the \textbf{\textit{SetB}} (like Liv, Kid, Sto, Pro) are clinically applicable, as these results are comparable or better trend than recent clinical assessments~\cite{liao2023comprehensive}. However, for complex structures and small organs, there are still performance gaps between these SOTA methods and clinical requirements.\\
\textbf{Robustness of \textbf{\textit{SetB}} and \textbf{\textit{SetC}}.} We further studied these SOTA methods' robustness and generalization on the two subsets consisting of clinically challenging cases, where these SOTA methods were trained on the \textbf{\textit{SetA}} and then were applied for testing directly. These results showed that all SOTA methods go to bad performance for patients with surgery resection, indicating the two subsets are challenging for recent methods. In addition, in the comparison between Table~\ref{tab:surgery-without-missing} and Table~\ref{tab:surgery-with-missing}, it can be observed that these SOTA methods presented a more significant performance drop, such as nnUNet with a performance drop of 2.53\% and 6.1\% in the term DSC, respectively. And, most of the organs with performance drop are the whole or partially resected organs after surgery or their neighbouring organs, like Kid, Gal, Adr, Rec, Pro, SVes. The potential reason may be the surgery breaks the normal anatomical structure or contextual appearance (like intensity distribution) and further leads to a performance drop. Afterwards, we further investigated the SOTA methods' performance on surgery resection organs where these resected organs should be predicted as background rather than any organ. Table~\ref{tab:hall_ratio} presented the organ hallucination ratio of these SOTA methods reformatted by false positive rate~\cite{rickmann2023halos}. It shows that most SOTA methods can not distinguish the surgery-resected organ effectively and produce hallucination prediction, suggesting that this challenging task needs more effort to achieve clinically applicable performance. \\
\textbf{Cross-evaluation between different datasets. }To build the general challenging cases benchmark, we studied the robustness and generalization of SOTA methods across different datasets.  Due to the different labelled organ categories, we just reported the results of the same annotated organ categories in Table~\ref{tab:raos2others} and Table~\ref{tab:cross}. Table~\ref{tab:raos2others} showed that there are domain gaps between RAOS and BTCV~\cite{landman2015miccai}, AbdomenCT-1K~\cite{ma2021abdomenct} and the domain shift between RAOS and BTCV is more significant than AbdomenCT-1K. Table~\ref{tab:cross} presented the cross-evaluation results between RAOS and AMOS showing that the domain gap is significant except for the without surgery subset. The above results further provided a new baseline and dataset for domain adaptation research.

\section{Conclusion}
In this study, we retrospectively collected 413 CT scans with 19 organ annotations to build an abdominal organ segmentation robust evaluation benchmark (RAOS). The RAOS dataset was split into three subsets according to patients' treatment strategies, without surgery, surgery without an organ missing and surgery with an organ missing. To the best of our knowledge, the RAOS is the first dataset for clinically relevant corner case segmentation evaluation. In addition, this work conducted a comprehensive evaluation of seven recent state-of-the-art methods and further investigated the domain gap between the proposed dataset and several public datasets (BTCV, AMOS, AbdomenCT-1K). The results showed that most recent works can not perform well when segmenting clinically challenging cases with irregular anatomy, such as patients with surgery. Meanwhile, it can be found that there are domain shifts between different datasets, and the RAOS can play a new role in the abdominal organ segmentation topic research. In the future, we will extend RAOS by adding more clinically challenging cases to boost the development of the clinically acceptable segmentation method.

\section{Acknowledgment} This work was supported by the National Natural Science Foundations of China [82203197, 62271115]. Note that, a small part of the RAOS dataset is from WORD\cite{luo2022word} (where these cases' annotations are extended from 16 classes to 19 classes).  In addition, we further extended the dataset with synthesised MR images. Please check the GitHub repo link.

\bibliographystyle{splncs04}
\bibliography{ref.bib}

\end{document}